\documentclass[11pt]{article}
\usepackage{amsmath,amssymb,color,epsfig,cite}
\usepackage{graphicx}
\usepackage{setspace}
\usepackage{mathrsfs}
\usepackage[english]{babel}
\usepackage{inputenc}
\usepackage{amsfonts,amsmath,amssymb}
\usepackage{mathrsfs}
\usepackage{graphicx}
\usepackage{color}
\usepackage{soul} 
\usepackage{hyperref}
\usepackage{subfigure}
\usepackage[normalem]{ulem}
\usepackage{array}
\usepackage{authblk}

\usepackage{amsfonts}
\newcommand{\hoch}[1]{$\, ^{#1}$}

\makeatletter
\@addtoreset{equation}{section}
\makeatother
\renewcommand{\theequation}{\thesection.\arabic{equation}}

\newcommand{\be}{\begin{equation}}
\newcommand{\ee}{\end{equation}}
\newcommand{\bea}{\setlength\arraycolsep{2pt} \begin{eqnarray}}
\newcommand{\eea}{\end{eqnarray}}

\def\0{{\sst{(0)}}}
\def\1{{\sst{(1)}}}
\def\2{{\sst{(2)}}}
\def\3{{\sst{(3)}}}
\def\4{{\sst{(4)}}}
\def\5{{\sst{(5)}}}
\def\6{{\sst{(6)}}}
\def\7{{\sst{(7)}}}
\def\8{{\sst{(8)}}}
\def\sst#1{{\scriptscriptstyle #1}}

\def\R{{\mathbb R}}
\def\C{{\mathbb C}}

\begin{document}

\vspace{15pt}
{\Large {\bf On the Nature of Bondi-Metzner-Sachs Transformations} }

\vspace{15pt}
{\bf Zahra Mirzaiyan\hoch{1} and Giampiero Esposito\hoch{2,1}}

\vspace{10pt}

\hoch{1} {\it INFN Sezione di Napoli, \\
Complesso Universitario di Monte S. Angelo, Via Cintia Edificio 6, 80126 Napoli, Italy.}
\hoch{2}{\it Dipartimento di Fisica ``Ettore Pancini'', \\
Complesso Universitario di Monte S. Angelo, Via Cintia Edificio 6, 80126 Napoli, Italy.}

\vspace{20pt}

\underline{ABSTRACT}

This paper investigates, as a first step, 
the four branches of BMS transformations, motivated by the
classification into elliptic, parabolic, hyperbolic and loxodromic proposed a
few years ago in the literature. We first prove that to each normal elliptic 
transformation of the complex variable $\zeta$ used in the metric for cuts 
of null infinity there corresponds a BMS supertranslation. We then study the conformal
factor in the BMS transformation of the $u$ variable as a function of the
squared modulus of $\zeta$. In the loxodromic and hyperbolic cases, this conformal
factor is either monotonically increasing or monotonically decreasing as a
function of the real variable given by the modulus of $\zeta$.
The Killing vector field of the Bondi metric is also studied in 
correspondence with the four admissible families of BMS transformations.
Eventually, all BMS transformations are re-expressed in the homogeneous
coordinates suggested by projective geometry. It is then found that BMS
transformations are the restriction to a pair of unit circles of a
more general set of transformations. Within this broader framework, the
geometry of such transformations is studied by means of its Segre manifold.
\vskip 1cm
\leftline{Type of the Paper: Article}

\noindent

\thispagestyle{empty}

\vfill
zahra.mirzaiyan@na.infn.it, gesposito@na.infn.it.

\pagebreak
\section{Introduction}
\setcounter{equation}{0}

The recent developments on the applications of the Bondi-Metzner-Sachs
(hereafter BMS) group, i.e., the asymptotic symmetry group of an
asymptotically flat spacetime (see Eqs. (A.1)-(A.3) of the Appendix),
have been motivated by black hole physics, quantum
gravity and gauge theories, as is well described in many outstanding works (e.g. Refs.
\cite{BMS01,BMS02,BMS03,BMS04,BMS05,BMS06,BMS07,BMS08,BMS09,BMS10,
BMS11,Donnay,Chowdhury,Bagchi,Compere}.
However, a purely classical investigation may still lead to a neater understanding
of the mathematical operations frequently performed. Within this framework,
at least four properties can be mentioned in our opinion:
\vskip 0.3cm
\noindent
(i) The proof by F. Alessio and one of us \cite{BMS12} that the BMS group is 
the right semidirect product of the proper orthocronous Lorentz group 
${\rm SO}^{+}(3,1)$ with supertranslations (cf. Appendix A).
\vskip 0.3cm
\noindent
(ii) The division of BMS transformations into parabolic, elliptic,
hyperbolic and loxodromic, since the first half of them consists of fractional
linear maps which can be classified by studying their fixed points \cite{BMS13}.
\vskip 0.3cm
\noindent
(iii) The investigation of fractional linear maps in general relativity
and quantum mechanics performed in Ref. \cite{Bellino}
\vskip 0.3cm
\noindent
(iv) The proof that the BMS group is not real analytic, and the related
suggestion that it is not locally exponential \cite{BMS14}.
\vskip 0.3cm
\noindent
(v) The recent discovery that groups of BMS type arise not only as
macroscopic asymptotic symmetry groups in cosmology, but describe 
also a fundamental microscopic symmetry of pseudo-Riemannian geometry
\cite{BMS15}.

In the following sections, we aim at presenting a detailed investigation
of the four branches of the BMS group and of yet other properties.
For this purpose, Sec. $2$ defines our basic framework, Sec. $3$ 
obtains a new theorem on supertranslations, Sec. $4$ studies the
conformal factor in the second half of BMS transformations, while 
Sec. $5$ studies Killing vector fields of the Bondi metric
and their behaviour under BMS transformations, obtaining 
a novel classification.
Eventually, BMS transformations in homogeneous coordinates are
studied in Sec. $6$, concluding remarks are presented in Sec. $7$,
while relevant details are provided in the Appendices.
The reader is referred to
Refs. \cite{BMS16,BMS17,BMS18} for the basic concepts of causal and
asymptotic structure of a spacetime manifold.

\section{Basic framework}
\setcounter{equation}{0}

The cuts of null infinity are spacelike $2$-surfaces orthogonal to the
generators of null infinity. Lengths within a cut scale by a variable factor $K$
under holomorphic bijections of the $2$-sphere $S^{2}$ to itself (hereafter we
use the complex variable $\zeta=e^{i \phi} \cot {\theta \over 2}$, 
$\phi$ and $\theta$ being the standard coordinates for $S^{2}$):
\begin{equation}
\zeta'=f(\zeta)={(a \zeta+b)\over (c \zeta +d)}
=f_{\Lambda}(\zeta),
\label{(2.1)}
\end{equation}
where 
$$
\Lambda=\left(\begin{matrix}
a & b \cr c & d 
\end{matrix}\right).
$$ 
This is a fractional linear (or M\"{o}bius) 
map, and the matrix $\Lambda$ can be
always taken to belong to ${\rm SL}(2,\C)$, because the ratio is
unaffected by rescalings of $a,b,c,d$ by the same factor, so that 
the passage from ${\rm GL}(2,\C)$ to ${\rm SL}(2,\C)$ is eventually
achieved. In particular, $f_{\Lambda}(\zeta)=f_{-\Lambda}(\zeta)$, 
because
$$
{(a\zeta+b)\over (c\zeta+d)}={(-a\zeta-b)\over (-c\zeta-d)}.
$$
Thus, two matrices of ${\rm SL}(2,\C)$ yield the same fractional
linear map if and only if the one is the opposite of the other.
At this stage, we are actually dealing with the projective version 
of the ${\rm SL}(2,\C)$ group, i.e. 
\begin{eqnarray}
{\rm PSL}(2,\C)&=& \left \{ 
(f,\Lambda)| \; 
f: \zeta \in \C \rightarrow 
f(\zeta)={(a\zeta+b)\over (c\zeta+d)}, \; ad-bc=1 \right \}
\nonumber \\
&=& {\rm SL}(2,\C)/\delta,
\label{(2.2)}
\end{eqnarray}
where $\delta$ is the homeomorphism such that
\begin{equation}
\delta(a,b,c,d)=(-a,-b,-c,-d).
\label{(2.3)}
\end{equation}
Our definition (2.2) of ${\rm PSL}(2,\C)$ as a space of maps
is formally analogous to the definition of ${\rm PSL}(2,\R)$ by
S. Katok \cite{BMS19}, and it puts the emphasis on the fractional
linear map associated to any matrix of ${\rm SL}(2,\C)$. Such maps
can be extended to the whole complex plane by defining \cite{BMS14}
\begin{equation}
f(\infty)={a \over c}, \;
f \left(-{d \over c}\right)=\infty.
\label{(2.4)}
\end{equation}
By virtue of the above considerations, we can consider the
equivalence relation
$$
(f_{\Lambda},\Lambda) \sim (f_{\Lambda'},\Lambda') 
\Longleftrightarrow \Lambda'= \pm \Lambda 
\Longrightarrow f_{\Lambda'}(\zeta)=f_{\Lambda}(\zeta).
$$

A cut remains the unit $2$-sphere under $f_{\Lambda}(\zeta)$ provided
that its metric is subject to the conformal rescaling 
\begin{equation}
K^{2}(\Lambda,\zeta) (d\theta \otimes d\theta +
\sin^{2}\theta d\phi \otimes d\phi),
\end{equation}
having defined
\begin{equation}
K(\Lambda,\zeta)=K_{\Lambda}(\zeta)={{1+|\zeta|^{2}}\over 
|a\zeta+b|^{2}+|c\zeta+d|^{2}},
\label{(2.6)}
\end{equation}
where of course $|\gamma|^{2}=\gamma {\overline \gamma}$ for all
$\gamma \in \C$. The asymptotic theory remains invariant under
this rescaling provided that lengths along the generators of null
infinity scale by the same amount, i.e.
\begin{equation}
du' = K_{\Lambda}(\zeta)du,
\label{(2.7)}
\end{equation}
which can be integrated to find
\begin{equation}
u'=K_{\Lambda}(\zeta)\Bigr[u+\alpha(\zeta,{\overline \zeta})\Bigr],
\label{(2.8)}
\end{equation}
where $\alpha$ is a suitably smooth function of $\zeta$ and of its
complex conjugate ${\overline \zeta}$. The transformations (2.1)
and (2.8) are related in such a way that they define the group of
BMS transformations
\begin{equation}
T(\zeta)=f_{\Lambda}(\zeta)={(a\zeta+b)\over (c\zeta+d)},
\label{(2.9)}
\end{equation}
\begin{equation}
T(u)=K_{\Lambda}(\zeta)\Bigr[u+\alpha(\zeta,{\overline \zeta})\Bigr].
\label{(2.10)}
\end{equation}
In a concise form, one can write \cite{BMS12}
\begin{equation}
T(\zeta,u)=(T(\zeta),T(u))=\Bigr(f_{\Lambda}(\zeta),
K_{\Lambda}(\zeta)\Bigr[u+\alpha(\zeta,{\overline \zeta}\Bigr]\Bigr).
\label{(2.11)}
\end{equation}
As pointed out in Ref. \cite{BMS13}, the transformations (2.9) 
can be classified according to their fixed points, for which 
$f(\zeta)=\zeta$. Hence only four families of fractional linear
maps are found to exist
\vskip 0.3cm
\noindent
(i) {\bf Parabolic}. Only one fixed point exists, for which
$(a+d)^{2}=4$, while
\begin{equation}
\Lambda=A_{P}=\left(\begin{matrix}
\pm 1 & \beta \cr
0 & \pm 1
\end{matrix}\right),
\label{(2.12)}
\end{equation}
\begin{equation}
f_{\Lambda}(\zeta)=f_{P}(\zeta)=\zeta \pm \beta.
\label{(2.13)}
\end{equation}
\vskip 0.3cm
\noindent
(ii) {\bf Elliptic}. Two fixed points exist, for which 
$(a+d)^{2}<4$, while
\begin{equation}
\Lambda=A_{E}=\left(\begin{matrix}
e^{i {\chi \over 2}} & 0 \cr
0 & e^{-i {\chi \over 2}}
\end{matrix}\right),
\label{(2.14)}
\end{equation}
\begin{equation}
f_{\Lambda}(\zeta)=f_{E}(\zeta)=e^{i \chi}\zeta.
\label{(2.15)}
\end{equation}
\vskip 0.3cm
\noindent
(iii) {\bf Hyperbolic}. Two fixed points exist, for which
$(a+d)^{2}>4$, while
\begin{equation}
\Lambda=A_{H}=\left(\begin{matrix}
\sqrt{|\kappa|} & 0 \cr
0 & {1 \over \sqrt{|\kappa|}}
\end{matrix}\right),
\label{(2.16)}
\end{equation}
\begin{equation}
f_{\Lambda}(\zeta)=f_{H}(\zeta)=|\kappa|\zeta.
\label{(2.17)}
\end{equation}
\vskip 0.3cm
\noindent
(iv) {\bf Loxodromic}. Two fixed points exist, for which
$(a+d)^{2} \in \C- \R$, and 
$$
(a+d)=\sqrt{k}+{1 \over \sqrt{k}}, \; 
k=\rho e^{i \sigma}, \; \rho \not = 1,
$$
while
\begin{equation}
\Lambda=A_{L}=\left(\begin{matrix}
\sqrt{\rho}e^{i{\sigma \over 2}} & 0 \cr
0 & {1 \over \sqrt{\rho}}e^{-i{\sigma \over 2}}
\end{matrix}\right),
\label{(2.18)}
\end{equation}
\begin{equation}
f_{\Lambda}(\zeta)=f_{L}(\zeta)=\rho e^{i \sigma} \zeta.
\label{(2.19)}
\end{equation}
Note that our matrices (2.12), (2.14), (2.16) and (2.18) belong
to ${\rm SL}(2,\C)$, whereas in section $2$ of Ref. \cite{BMS13} 
only $A_{P}=M_{P}$ was in ${\rm SL}(2,\C)$, whereas the matrices
$M_{E},M_{H}$ and $M_{L}$ therein were elements of ${\rm GL}(2,\C)$. 

Since also the transformation $T(u)$ depends on the matrix $\Lambda$
through the conformal factor $K_{\Lambda}(\zeta)$, the work in Ref.
\cite{BMS13} proposed the same nomenclature, from parabolic to 
loxodromic, for the whole group of BMS transformations in Eq. (2.11).
By virtue of Eqs. (2.6) and (2.12), (2.14), (2.16) and (2.18) 
one finds therefore
\begin{equation}
K_{P}(\zeta)={{1+|\zeta|^{2}}\over
\Bigr(1+|\pm \beta + \zeta|^{2}\Bigr)},
\label{(2.20)}
\end{equation}
\begin{equation}
K_{E}(\zeta)=1,
\label{(2.21)}
\end{equation}
\begin{equation}
K_{H}(\zeta)= {|\kappa|(1+|\zeta|^{2}) \over \Bigr(1+|\kappa|^{2}
|\zeta|^{2}\Bigr)},
\label{(2.22)}
\end{equation}
\begin{equation}
K_{L}(\zeta)={\rho(1+|\zeta|^{2}) \over \Bigr(1+\rho^{2}
|\zeta|^{2}\Bigr)},
\label{(2.23)}
\end{equation}
in the parabolic, elliptic, hyperbolic and loxodromic cases,
respectively. Once more, our Eqs. (2.22) and (2.23) differ by a
multiplicative factor from the Eqs. in section $4$ of Ref.
\cite{BMS13} because all our matrices are in ${\rm SL}(2,\C)$.

\section{A new theorem on supertranslations}
\setcounter{equation}{0}

At this stage, we can immediately prove the following theorem:
\vskip 0.3cm
\noindent
{\bf T1}. A normal elliptic transformation, where the phase factor
$\chi$ in Eq. (2.15) is an integer multiple of $2\pi$, engenders a
BMS supertranslation.
\vskip 0.3cm
\noindent
{\bf Proof}. If $\chi=2\pi l$, $l$ being a relative integer, one finds
the BMS transformation
$$
T(\zeta)=f_{E}(\zeta)=\zeta,
$$
which implies that
\begin{equation}
T(\theta)=\theta, \; T(\phi)=\phi,
\label{(3.1)}
\end{equation}
as well as (see Eqs. (2.10) and (2.21))
\begin{equation}
K_{E}(\zeta)=1 \Longrightarrow T(u)=
u+\alpha(\zeta,{\overline \zeta}).
\label{(3.2)}
\end{equation}
Equations (3.1) and (3.2) are precisely the defining equations
of the Abelian subgroup of supertranslations \cite{BMS12}. In
other words, restriction to normal elliptic transformations, 
jointly with a choice of the function $\alpha$, engenders all
supertranslations. Q.E.D.\\
As an explicit example, let us consider the most general metric in four dimensions in 
Bondi coordinates $(u,r,\zeta, {\overline \zeta})$:
\begin{equation}
g=- U du \otimes du-e^\beta (du \otimes dr+dr \otimes du)+g_{AB} (dx^A+\frac{1}{2} U^A du)  
\otimes (dx^B+\frac{1}{2} U^B du),
\label{(3.3)}
\end{equation}
where $x^A=(\zeta, {\overline \zeta})$. The local diffeomorphism invariance 
is fixed by the following conditions:
\begin{equation}
\partial_r \text{det} \Bigr(\frac{g_{AB}}{r^2}\Bigr)=0, \ \ \ \ g_{rr}=g_{rA}=0.
\label{(3.4)}
\end{equation}
In order to eliminate six Lorentz generators and thereby eliminating boosts and rotations that grow with $r$ 
at infinity, we restrict ourselves to the diffeomorphisms generated by the vector field 
$\varepsilon$ whose components have the large-$r$ falloffs:
\begin{equation}
{ }^{(u)}\varepsilon, \; { }^{(r)}\varepsilon \sim 
\mathcal{O}(r^0),\ \ \ { }^{(\zeta)}\varepsilon, 
{ }^{({\overline \zeta})}\varepsilon 
\sim \mathcal{O}(\frac{1}{r}).
\label{(3.5)}
\end{equation}
By definition, the asymptotic symmetries must preserve the falloff conditions:
\begin{eqnarray}
&&g_{uu}=-1+\frac{2m_B}{r}+\mathcal{O}(r^{-2}),\nonumber\\
&&g_{ur}=-1+\mathcal{O}(r^{-2}),\nonumber\\
&&g_{uA}=\frac{1}{2} D^B C_{BA}+\mathcal{O}(r^{-1}),\nonumber\\
&&g_{AB}=r^2 \gamma_{AB}+r C_{AB}+\mathcal{O}(r^0),
\label{(3.6)}
\end{eqnarray}
where $m_B$ is known as the Bondi mass aspect and $C_{\zeta\zeta}$ describes gravitational waves at 
large $r$. Moreover, $\gamma_{AB}$ is the metric on the 2-sphere described by 
\begin{eqnarray}
\gamma_{\zeta\overline \zeta}=\frac{2}{(1+\zeta \overline \zeta)^2}.
\label{(3.7)}
\end{eqnarray}
By using the falloff conditions (\ref{(3.5)}), one finds in Bondi gauge
\begin{eqnarray}
(\mathcal{L}_{\varepsilon} g)_{rr}=2 g_{ur} \partial_r { }^{(u)}\varepsilon ,
\label{(3.8)}
\end{eqnarray}
which implies that ${ }^{(u)}\varepsilon$ must be independent of $r$. In addition, to the 
leading order in the asymptotic expansion
\begin{eqnarray}
(\mathcal{L}_{\varepsilon} g)_{ur}= g_{ur} \partial_u { }^{(u)}\varepsilon 
+\mathcal{O}(r^{-1}).
\label{(3.9)}
\end{eqnarray}
Hence, ${ }^{(u)}\varepsilon(\zeta, \overline \zeta)$ 
is a function on the 2-sphere which we fix as being equal to 
$\alpha(\zeta, \overline\zeta)$. Then, requiring that the Bondi conditions (\ref{(3.4)}) 
and the falloffs (\ref{(3.6)}) are preserved implies that at large $r$ 
\begin{eqnarray}
\varepsilon=\alpha(\zeta, \overline \zeta)\  \partial_u+D^\zeta D_\zeta\  \alpha(\zeta, \overline \zeta) \ 
\partial_r-\frac{1}{r}D^\zeta \ \alpha(\zeta, \overline \zeta)\ \partial_\zeta+ c.c+...,
\label{(3.10)}
\end{eqnarray}
where $\alpha(\zeta, \overline \zeta)$ can be any function of $(\zeta, \overline \zeta)$ on 
the 2-sphere. The function $\alpha(\zeta,{\overline \zeta})$ can be expanded in spherical harmonics 
on the 2-sphere.  The modes $l=0$ and $l=1$ correspond to the standard global translations in Minkowski 
space-time. The vector field $\varepsilon (\alpha(\zeta,{\overline \zeta}))$ on the $l=0$ 
and $l=1$ spherical harmonics can be evaluated as 
\begin{eqnarray}
&&\varepsilon (Y_0^0)=Y_0^0\ \partial_u,\nonumber\\
&&\varepsilon (Y_1^m)=Y_1^m\ \partial_u+\frac{1}{2} D^2 Y_1^m \partial_r
-\frac{\gamma^{AB}\ \partial_B Y_1^m}{r} \ \partial_A,
\label{(3.11)}
\end{eqnarray}
with the following normalization for spherical harmonics:
\begin{eqnarray}
Y_0^0=1, \ \ Y_1^1= \frac{\zeta}{(1+\zeta {\overline \zeta})}, \ \ Y_1^0= \frac{(1-\zeta {\overline \zeta})}
{(1+\zeta {\overline \zeta})}, \ \ Y_1^{-1}= \frac{\overline\zeta}{(1+\zeta {\overline \zeta})}.
\label{(3.12)}
\end{eqnarray}
Then the standard global translations in Minkowski space-time are defined as 
\begin{eqnarray}
&&\varepsilon(Y_0^0)=\partial_u,\nonumber\\
&&\varepsilon(Y_1^1)=\frac{\zeta}{(1+\zeta {\overline \zeta})} \ (\partial_u-\partial_r)+\frac{\zeta^2}{2r} 
\partial_\zeta-\frac{1}{2r}\partial_{\overline\zeta},\nonumber\\
&&\varepsilon(Y_1^0)=\frac{(1-\zeta {\overline \zeta})}{(1+\zeta {\overline \zeta})} \ (\partial_u-\partial_r)
+\frac{\zeta}{r} \partial_\zeta-\frac{\overline\zeta}{r}\partial_{\overline\zeta},\nonumber\\
&&\varepsilon(Y_1^{-1})=\frac{{\overline \zeta}}{(1+\zeta {\overline \zeta})} \ (\partial_u-\partial_r)+\frac{1}{2r} 
\partial_\zeta-\frac{\overline\zeta^2}{2r}\partial_{\overline\zeta}.
\label{(3.13)}
\end{eqnarray}
Other choices of $l$ engender all supertranslations.

\section{Behaviour of the conformal factor}
\setcounter{equation}{0}

The conformal factors (2.20), (2.22) and (2.23) have the limiting
behaviours 
\begin{equation}
\lim_{|\zeta| \to 0}K_{P}={1 \over (1+|\beta|^{2})}, \;
\lim_{|\zeta| \to \infty}K_{P}=1,
\label{(4.1)}
\end{equation}
\begin{equation}
\lim_{|\zeta| \to 0}K_{H}=|\kappa|, \;
\lim_{|\zeta| \to \infty}K_{H}={1 \over |\kappa|},
\label{(4.2)}
\end{equation}
\begin{equation}
\lim_{|\zeta| \to 0}K_{L}=\rho, \;
\lim_{|\zeta| \to \infty}={1 \over \rho}.
\label{(4.3)}
\end{equation} 
Moreover, since the independent variable $x=|\zeta|$ is always 
$\geq 0$, both $K_{H}$ and $K_{L}$ can be studied by 
considering the function
\begin{equation}
F: x \in [0,\infty] \rightarrow F(x)={\xi (1+x^{2}) \over
(1+\xi^{2}x^{2})},
\label{(4.4)}
\end{equation}
where $\xi=|\kappa|$ or $\xi=\rho$ in the hyperbolic and 
loxodromic cases, respectively. Since the first two
derivatives of $F$ are given by
\begin{equation}
F'(x)={2\xi (1-\xi^{2})x \over (1+\xi^{2}x^{2})^{2}}, \;
F''(x)={2\xi(1-\xi^{2})\over (1+\xi^{2}x^{2})^{3}}
(1-3\xi^{2}x^{2}),
\label{(4.5)}
\end{equation}
we find that, if $\xi \in ]0,1[$, the function $F$ is monotonically
increasing $\forall x \in [0,\infty]$, displays an upwards concavity
and takes its absolute minimum at $x=0$. The figure below plots
the graph of $F$ when $\xi$ is either less than $1$ or bigger
than $1$. 

 \begin{figure}[!h]
\centering
\includegraphics[width=7.3cm]{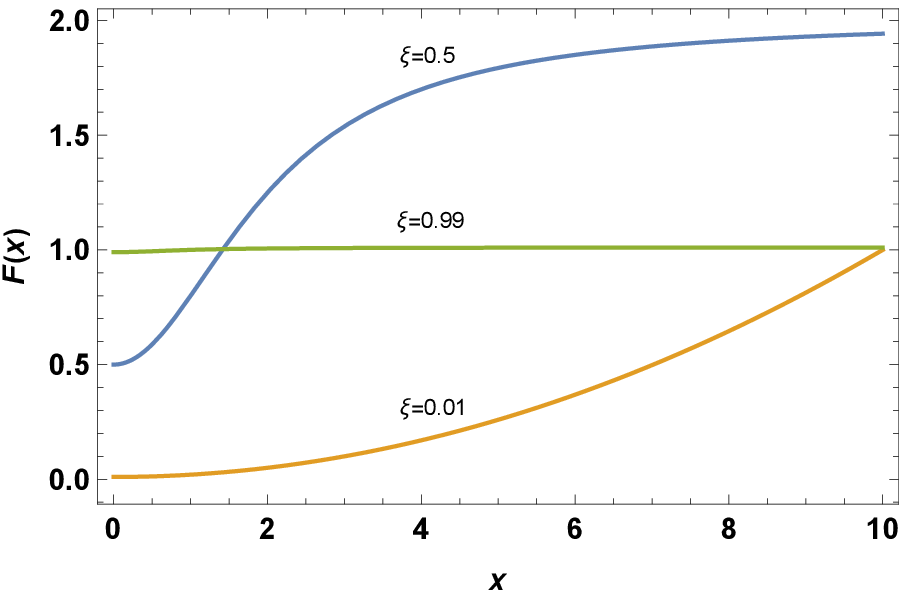}\hfill
\includegraphics[width=7.3cm]{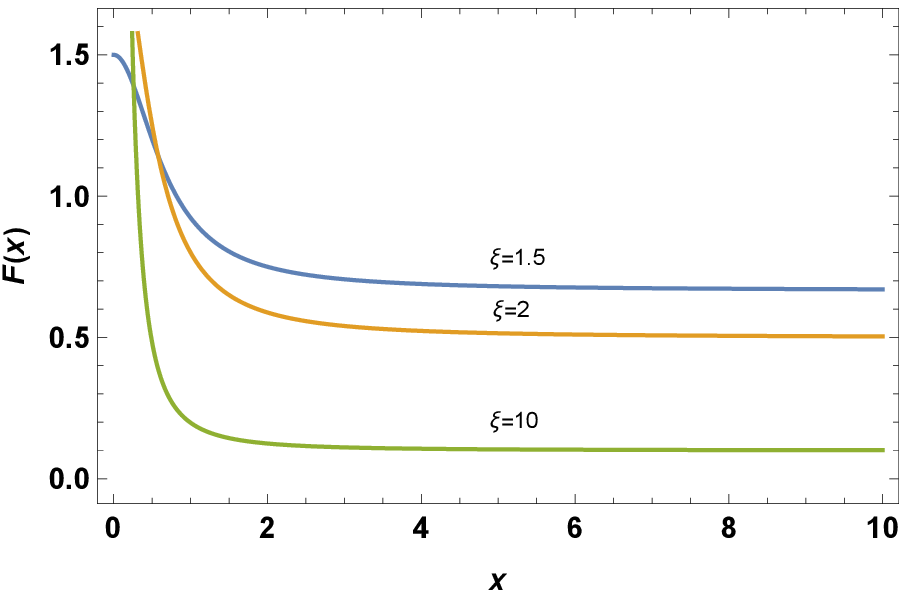}
\caption[] {{The conformal factor is monotonically increasing if $\xi \in ]0,1[$ and monotonically 
decreasing if $\xi >1$ in the hyperbolic and loxodromic cases.}}\label{figure:Fx}
 \end{figure}

In the parabolic case, the conformal factor given in Eq. (2.20) can
be re-expressed in the form
\begin{equation}
K_{P}(\zeta)={{1+|\zeta|^{2}}\over 
\left[1+|\beta|^{2}+\left(1 \pm 2 {\rm Re}\left({\beta \over \zeta}\right)
\right)|\zeta|^{2}\right]},
\label{(4.6)}
\end{equation}
and hence we cannot exploit the theory of functions of a
real variable for the parabolic conformal factor. The figure below plots the graph of the conformal 
factor $K_P(\zeta)$ in the $(\zeta, \overline \zeta)$ plane. Of course, using either Eq. (2.20) or Eq. 
(4.6) leads to the same plot  ($K_{P+}$ and $K_{P-}$ are devoted to the plus-minus in the denominator 
of the conformal factor $K_P (\zeta)$).

\begin{figure*}
\centering
\begin{tabular}{ccc}
\rotatebox{0}{
\includegraphics[width=0.5\textwidth,height=0.27\textheight]{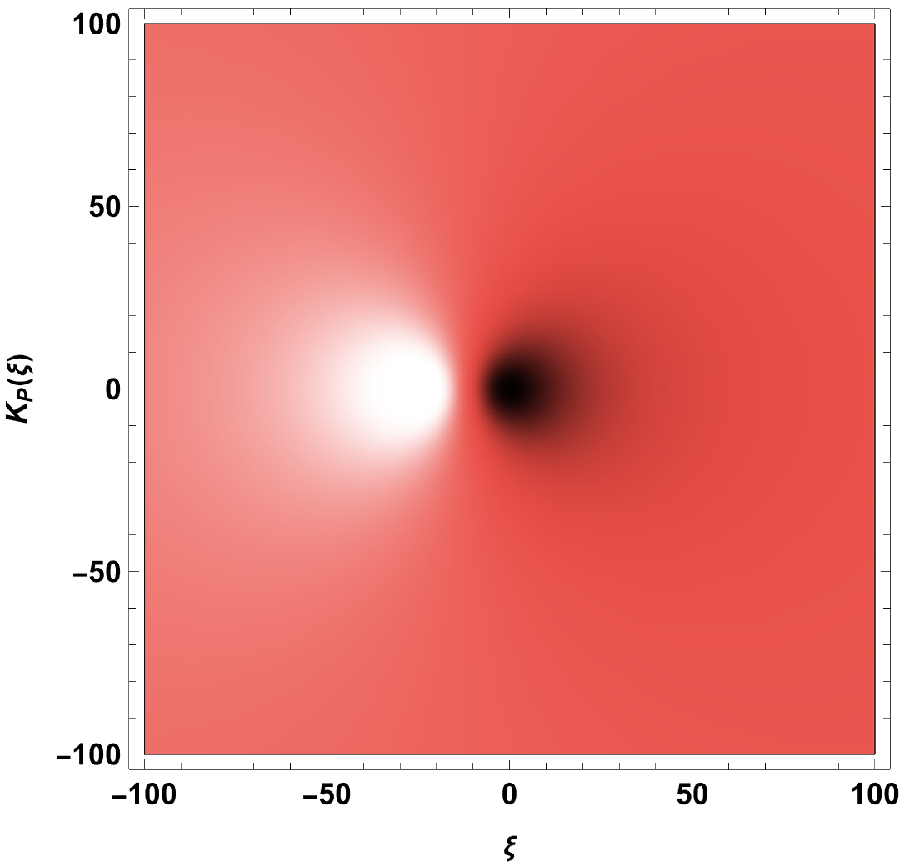}}&
\rotatebox{0}{
\includegraphics[width=0.5\textwidth,height=0.27\textheight]{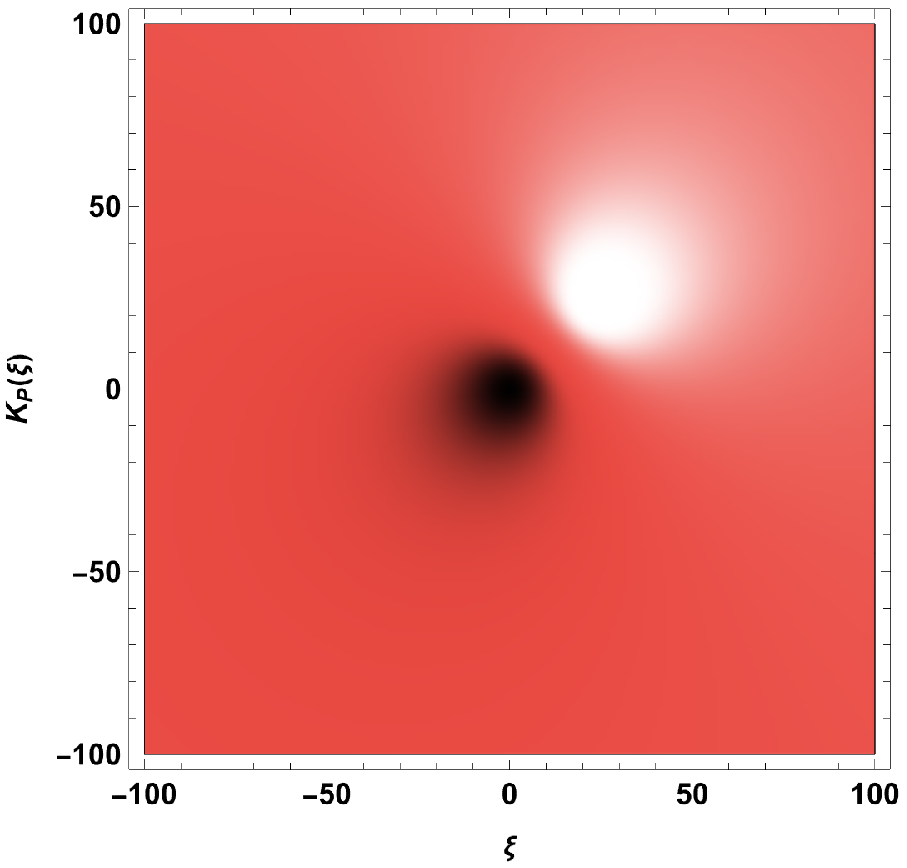}}\\
\rotatebox{0}{
\includegraphics[width=0.5\textwidth,height=0.27\textheight]{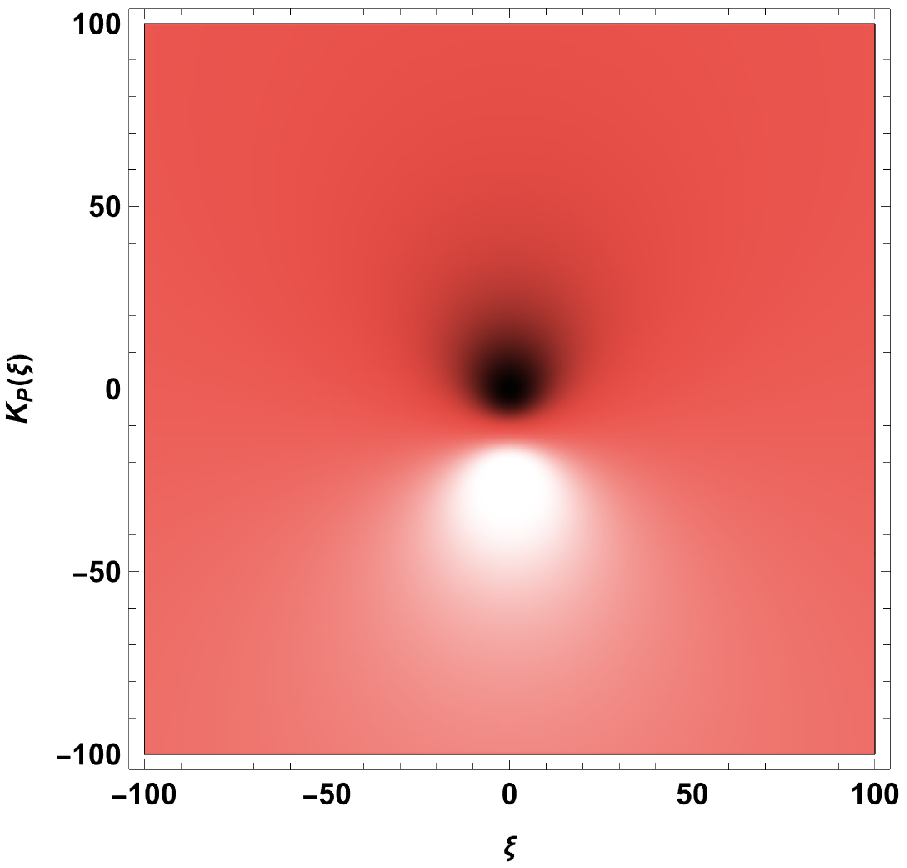}}&
\rotatebox{0}{
\includegraphics[width=0.5\textwidth,height=0.27\textheight]{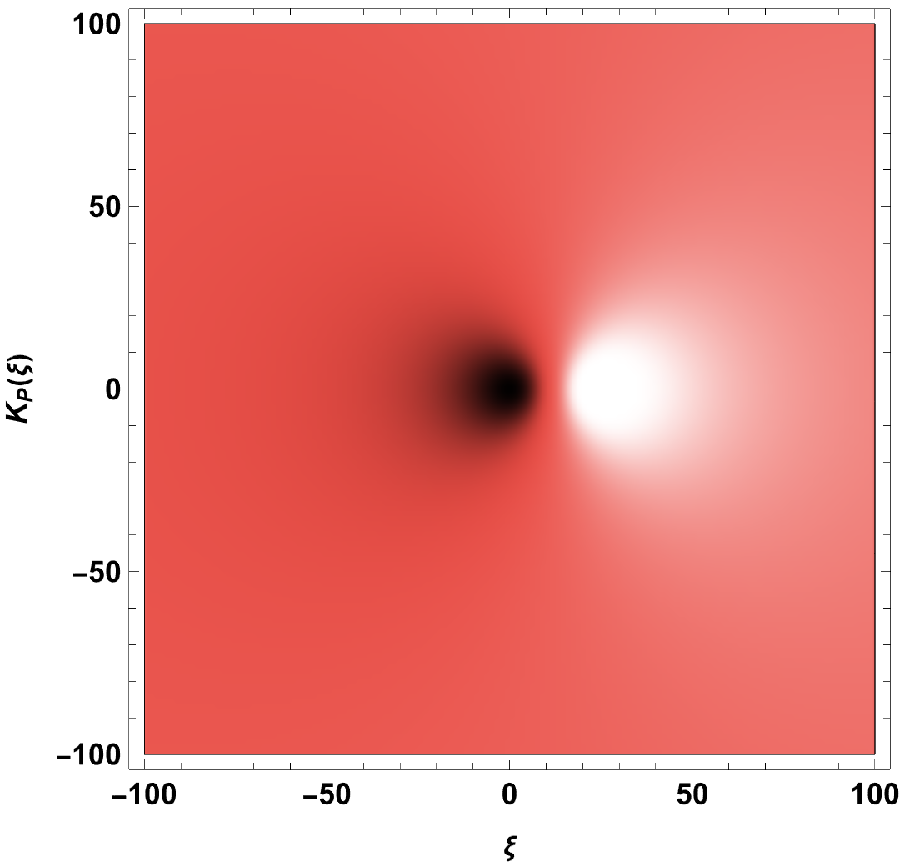}}\\
\rotatebox{0}{
\includegraphics[width=0.5\textwidth,height=0.27\textheight]{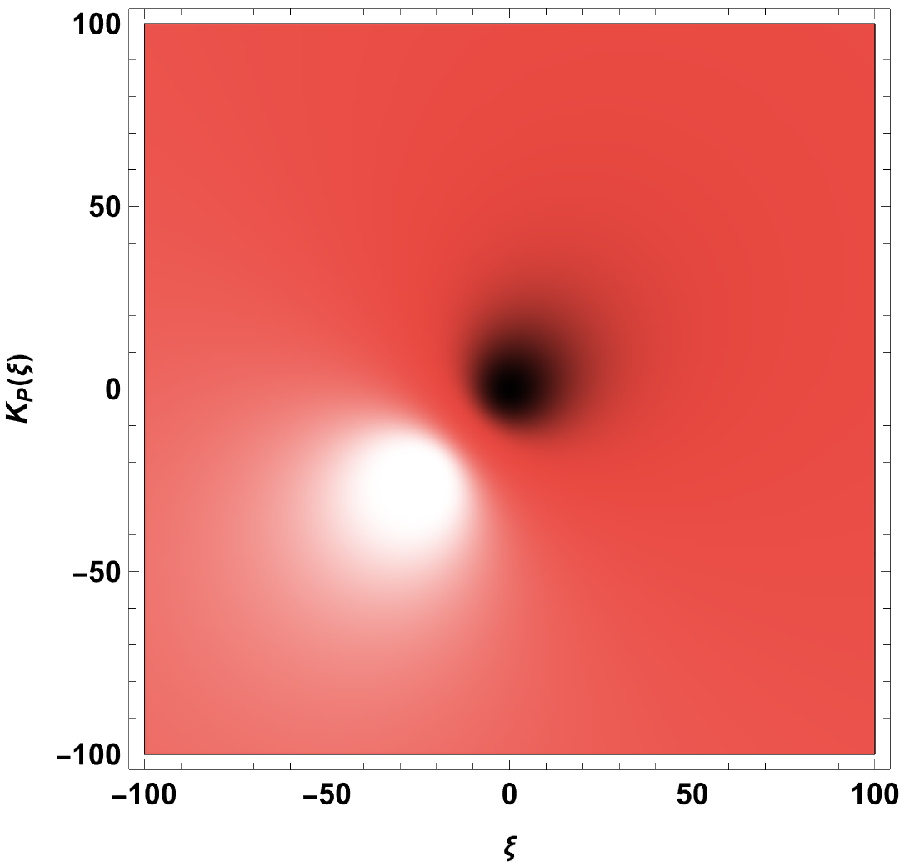}}&
\rotatebox{0}{
\includegraphics[width=0.5\textwidth,height=0.27\textheight]{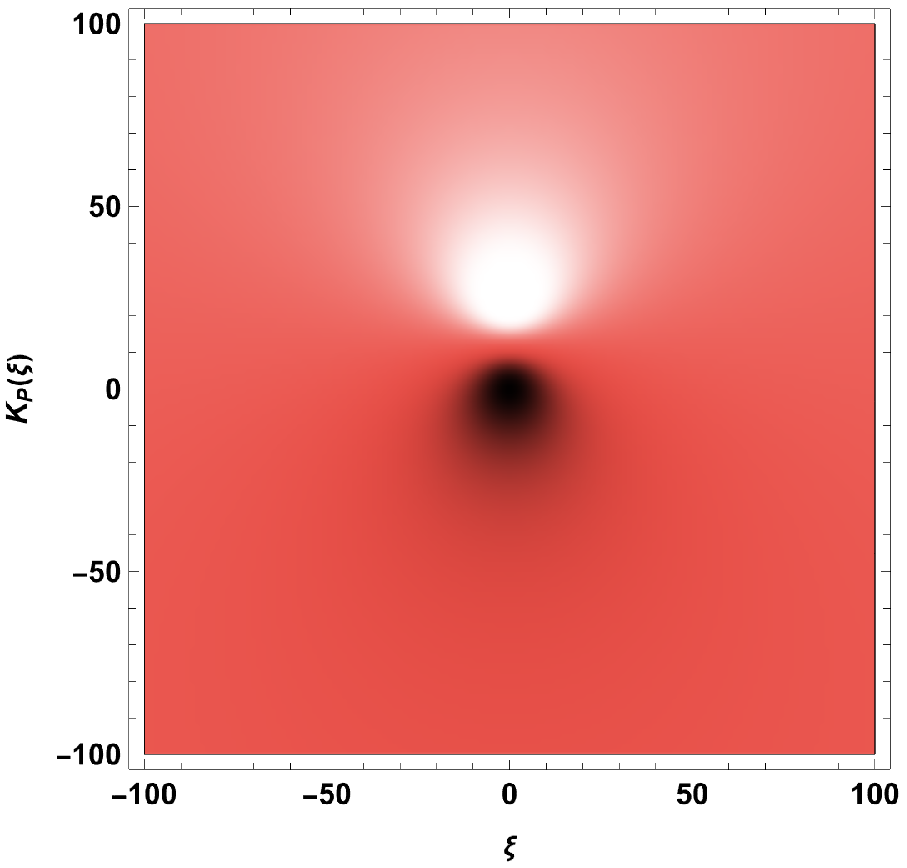}}\\
\\
\end{tabular}
\caption{$K_{P} (\zeta)$ in the $(\zeta, \overline\zeta)$ plane. First row from left to right: 
$K_{P+}$ with $\beta$ as a real parameter, $K_{P+}$ with $\beta$ as a complex parameter.
Second row:
$K_{P+}$ with $\beta$ as a purely imaginary parameter, $K_{P-}$ with $\beta$ as a real parameter.
Third row:
$K_{P-}$ with $\beta$ as a complex parameter, $K_{P-}$ with $\beta$ as a purely imaginary parameter.
}\label{figure:RE1}
\end{figure*}

\section{Behaviour of Killing vector fields under BMS transformations}
\setcounter{equation}{0}
It is interesting to derive the most general form of the diffeomorphism associated 
with the four branches of the BMS group. We look for a diffeomorphism $\epsilon$ which 
satisfies the asymptotic falloff condition defined in Eq. (\ref{(3.5)}) together with 
the asymptotic symmetries preserving the falloff conditions described in Eq. (3.6) in 
Bondi gauge. As already mentioned in Eqs. (\ref{(2.9)}) and (\ref{(2.10)}), the group of
BMS transformations is defined as 
\begin{eqnarray}
&&T(\zeta)=f_{\Lambda}(\zeta)={(a\zeta+b)\over (c\zeta+d)} \nonumber\\
&&T(u)=K_{\Lambda}(\zeta)\Bigr[u+\alpha(\zeta,{\overline \zeta})\Bigr].
\label{(5.1)}
\end{eqnarray}
We recall that the first line of Eq. (\ref{(5.1)}) 
can be always reduced to one of the forms (2.13), (2.15), (2.17)
or (2.19), where $f_{\Lambda}(\zeta)$ reads eventually
$$
f_{\Lambda}(\zeta)={\cal N}_{\gamma}\zeta+\gamma.
$$
Moreover, we consider the asymptotic expansion of the vector field $\varepsilon$
\begin{equation}
\varepsilon= { }^{(u)}\varepsilon \partial_u+\sum_{n=0}^\infty 
\frac{{ }^{(r)}\varepsilon_n}{r^n} 
\partial_r+\sum_{n=1}^\infty \frac{{ }^{(\zeta)}\varepsilon_n}{r^n} \partial_{\zeta}
+\sum_{n=1}^\infty \frac{{ }^{({\overline \zeta})}\varepsilon_n}{r^n} \partial_{\overline\zeta}.
\label{(5.2)}
\end{equation}
The variation of the metric under a diffeomorphism is given by
\begin{equation}
(\mathcal{L}_\varepsilon g)_{\mu\nu}=\varepsilon^\rho\partial_\rho g_{\mu\nu}
+g_{\mu\rho}\partial_\nu \varepsilon^\rho+g_{\nu\rho}\partial_\mu \varepsilon^\rho.
\label{(5.3)}
\end{equation}
In Bondi gauge, 
\begin{equation}
(\mathcal{L}_\varepsilon g)_{rr}=2 g_{ur} \partial_r { }^{(u)}\varepsilon ,
\label{(5.4)}
\end{equation}
which implies that ${ }^{(u)}\varepsilon$ must be independent of $r$:
\begin{equation}
{ }^{(u)}\varepsilon={ }^{(u)}\varepsilon (u,\zeta,\overline\zeta).
\label{(5.5)}
\end{equation}
By using Eq. (\ref{(2.7)}) together with the falloff conditions for the 
metric, the leading order in the asymptotic expansion gives
\begin{equation}
(\mathcal{L}_\varepsilon g)_{ur}=-\partial_u { }^{(u)}\varepsilon 
=- (K_\Lambda(\zeta) -1),
\label{(5.6)}
\end{equation}
where the last equality is obtained by evaluating the difference 
between the metric when $du$ is conformally rescaled according
to (2.7) and the original metric with no rescaling of $du$.
Equation (5.6) suggests the following form for ${ }^{(u)}\varepsilon$
\begin{equation}
{ }^{(u)}\varepsilon=F(\zeta,\overline\zeta)+u\ (K_\Lambda(\zeta)-1).
\label{(5.7)}
\end{equation}
From $(\mathcal{L}_\varepsilon g)_{r\zeta}$ at order $\mathcal{O}(r^0)$, one finds 
\begin{equation}
{ }^{(\zeta)}\varepsilon_1=- {\cal N}_\gamma\ D^\zeta { }^{(u)}\varepsilon.
\label{(5.8)}
\end{equation}
Moreover, $(\mathcal{L}_\varepsilon g)_{r\zeta}$ at order $\mathcal{O}(r^{-1})$ gives us 
\begin{equation}
{ }^{(\zeta)}\varepsilon_2=\frac{{\cal N}_\gamma}{2} C^{\zeta\zeta} 
D_\zeta { }^{(u)}\varepsilon .
\label{(5.9)}
\end{equation}
The leading order of $\mathcal{O}(r)$ term of $(\mathcal{L}_\varepsilon g)_{uu}$ requires 
\begin{equation}
{ }^{(r)}\varepsilon_0= G(\zeta,\overline\zeta) +{u\over 2} (K_\Lambda^2-1).
\label{(5.10)}
\end{equation}
The function $G(\zeta,\overline\zeta)$ can be defined from the traceless nature of 
the $\mathcal{O}(r)$ term of $(\mathcal{L}_\varepsilon g)_{\zeta\zeta}$ as
\begin{equation}
{ }^{(r)}\varepsilon_0=\frac{\mathcal{N}^2_\gamma}{2} D^2 
{ }^{(u)}\varepsilon +{u\over 2} (K_\Lambda^2-1).
\label{(5.11)}
\end{equation}
Thus, the Killing vector field of the Bondi metric in correspondence 
with the four branches of the BMS transformations reads as
\begin{eqnarray}
&&\varepsilon= \big(F(\zeta, \overline\zeta) +u (K_\Lambda(\zeta)-1)\big) \partial_u
+ \left(\frac{{\cal N}^2_\gamma}{2} D^2 { }^{(u)}\varepsilon
+{u\over 2} (K_\Lambda^2-1)\right)\partial_r
\nonumber\\
&&-\frac{1}{r} \big({\cal N}_\gamma\ D^\zeta (F(\zeta, \overline\zeta)
+u\ (K_\Lambda(\zeta)-1))\big)\partial_\zeta+c.c+...
\label{(5.12)}
\end{eqnarray}
Hence four families of diffeomorphisms in correspondence with the BMS transformations exist:\\
(i) {\bf Parabolic}. In the case of a parabolic fractional linear map for $\zeta$
\begin{eqnarray}
&&K_{P}(\zeta)={{1+|\zeta|^{2}}\over
\Bigr(1+|\pm \beta + \zeta|^{2}\Bigr)},\\
&&{\cal N}_{\gamma}=
{\cal N}_P=\pm 1, \ \ \ 
\gamma=\gamma_{P}=\beta.
\label{(5.13)}
\end{eqnarray}
(ii) {\bf Elliptic}. For an elliptic fractional linear map,
\begin{eqnarray}
&&K_{E}(\zeta)=1,\\
&&{\cal N}_{\gamma}
={\cal N}_E=1, \ \ \ 
\gamma=\gamma_{E}=0.
\label{(5.15)}
\end{eqnarray} 
Therefore, the Killing vector field (\ref{(5.12)}) coincides with the form 
obtained in Eq. (\ref{(3.10)}).
(iii) {\bf Hyperbolic}.
\begin{eqnarray}
&&K_{H}(\zeta)= {|\kappa|(1+|\zeta|^{2}) \over \Bigr(1+|\kappa|^{2}
|\zeta|^{2}\Bigr)},\\
&&{\cal N}_\gamma={\cal N}_{H}=|k|, \ \ \ 
\gamma=\gamma_{H}=0.
\label{(5.17)}
\end{eqnarray} 
(iv) {\bf Loxodromic}.
\begin{eqnarray}
&&K_{L}(\zeta)={\rho(1+|\zeta|^{2}) \over \Bigr(1+\rho^{2}
|\zeta|^{2}\Bigr)},\\
&&{\cal N}_\gamma={\cal N}_{L}=\rho e^{i\sigma}, \ \ \ 
\gamma=\gamma_{L}=0.
\label{(5.19)}
\end{eqnarray} 
 
Thus, the Killing vector fields associated with the four branches of the 
BMS transformations have been here derived for the first time
in the literature by substituting  
$K_\Lambda(\zeta)$, ${\cal N}_\gamma$ and $\gamma$ 
from Eqs. (5.13)-(5.20) into Eq. (\ref{(5.12)}).

\section{BMS transformations in homogeneous coordinates}
\setcounter{equation}{0}

The material at the beginning of Appendix B suggests expressing 
our complex variable $\zeta=e^{i \phi}\cot {\theta \over 2}$ 
in the form $\zeta={z_{0}\over \zeta_{1}}$. This is easily 
accomplished by defining
\begin{equation}
z_{0} \equiv e^{i {\phi \over 2}} \cos {\theta \over 2}, \;
z_{1} \equiv e^{-i {\phi \over 2}} \sin {\theta \over 2},
\label{(6.1)}
\end{equation}
and hence writing the first half of BMS transformations, our 
Eq. (2.1), in the form (B.1):
\begin{equation}
\left(\begin{matrix}
z_{0}' \cr z_{1}'
\end{matrix}\right)
= \left(\begin{matrix}
a & b \cr
c & d 
\end{matrix}\right)
\left(\begin{matrix}
z_{0} \cr z_{1} 
\end{matrix}\right), 
\label{(6.2)}
\end{equation}
where the ${\rm SL}(2,\C)$ matrix can only be either (2.12), or
(2.14), or (2.16), or (2.18).

The second half of BMS transformations, our Eq. (2.8), now reads as
\begin{equation}
u'=K_{\Lambda}(z_{0},z_{1})\Bigr[u
+\alpha(z_{0},z_{1};{\bar z}_{0},{\bar z}_{1})\Bigr],
\label{(6.3)}
\end{equation}
where the conformal factor can only take one of the four forms 
(2.20)-(2.23), upon setting $\zeta={z_{0}\over z_{1}}$ therein.

In Eqs. (6.2) and (6.3), the complex variables are defined as
in (6.1) and obey therefore the restrictions
$$
|z_{0}| \leq 1, \; |z_{1}| \leq 1,
$$
i.e. they correspond to a pair of unit circles $\Gamma_{0}$ and
$\Gamma_{1}$. Thus, we may recognize that the BMS transformations 
are the restrictions to these circles of a more general set of
transformations, i.e.
\begin{equation}
\left(\begin{matrix}
w_{1}' \cr w_{2}' 
\end{matrix}\right)
= \left(\begin{matrix}
a & b \cr c & d 
\end{matrix}\right)
\left(\begin{matrix}
w_{1} \cr w_{2} \end{matrix}\right),
\label{(6.4)}
\end{equation}
\begin{equation}
u'=K_{\Lambda}(w_{1},w_{2}) 
\Bigr[u + \alpha(w_{1},w_{2};{\bar w}_{1},{\bar w}_{2}\Bigr],
\label{(6.5)}
\end{equation}
where both $w_{1}$ and $w_{2}$ belong to $\C \cup \left \{ \infty \right \}$,
and they are such that
$$
w_{1}|_{\Gamma_{0}}=z_{0}, \;
w_{2}|_{\Gamma_{1}}=z_{1}.
$$
Within this broader framework, one can consider two complex projective
planes. Let $P$ be a point of the first plane with coordinates 
$(w_{0},w_{1},w_{2})$, and let $P'$ be a point of the second plane,
with coordinates $(u_{0},u_{1},u_{2})$. We can now take all nine
products between a complex coordinate of $P$ and a complex coordinate
of $P'$, i.e.
\begin{equation}
Z_{hk}=w_{h}u_{k}, \; h,k=0,1,2.
\label{(6.6)}
\end{equation}
These nine complex quantities are defined up to a proportionality
factor, since this is the case for both $w_{h}$ and $u_{k}$. They 
can be therefore interpreted as the coordinates of a point $Z$ of
eight-dimensional complex projective space $S_{8}$. To the pair
of points $P$ and $P'$ there corresponds a point $Z$ of $S_{8}$ 
by means of Eq. (6.6). As $P$ and $P'$ are varying in their own
plane, the point $Z$ describes in $S_{8}$ a four-complex-dimensional 
manifold, since both $P$ and $P'$ are varying on a plane, i.e. 
a two-complex-dimensional geometric object. Equations (6.6) 
represent therefore a four-complex-dimensional manifold $V_{4}$
in the complex projective space $S_{8}$. Such a manifold is the
{\it Segre manifold} \cite{Caccioppoli,Beltrametti}.

If in the first plane we fix the point $P=(w_{0},w_{1},w_{2})$,
Eqs. (6.6) become linear homogeneous in the $u_{k}$ coordinates
and, as such, they represent a plane in $S_{8}$. Thus, to every
point of the first plane there corresponds a plane on the Segre
manifold $V_{4}$. The Segre manifold contains therefore a
complex double infinitude of planes. In completely analogous 
way, another double infinitude of planes of $V_{4}$ corresponds to 
the double infinitude of points of the second plane. A plane of
this second infinitude is obtained by fixing a point $P'$ in the
second plane and then letting $P$ vary in the first plane.
Each of these $\infty^{1}$ systems of planes is an array, in light
of the correspondence between elements of the system and points
of a plane. Hence the Segre manifold contains two arrays of
planes. Two planes of the same array do not have common points,
whereas two planes belonging to different arrays have one and
only one common point \cite{Caccioppoli}.

One can also fix the point $P$ and let the point $P'$ vary not
over the whole plane, but only on a line in such a plane. 
In correspondence one obtains on the Segre manifold $V_{4}$ a
$V_{1}$ subset, i.e. a curve. If both $P$ and $P'$ describe 
a line in their own plane, one obtains on the Segre manifold
$V_{4}$ a $V_{2}$ subset, i.e. a quadric. Hence to every pair
of lines there corresponds a quadric. Since there exist 
$\infty^{2}$ lines in a plane, the quadrics of a Segre manifold
are $\infty^{4}$. In other words, the Segre manifold contains
a complex fourfold infinity of quadrics.

At a deeper level, we can say that the Segre manifold 
is the projective image of the product of projective 
spaces, and it is a natural tool for studying the
framework where we can accommodate the transformations
that reduce to the BMS transformations upon restriction
to the pair of unit circles $\Gamma_{0}$ and $\Gamma_{1}$.

\section{Concluding remarks}
\setcounter{equation}{0}

Since asymptotic flatness is a limiting case of classical
general relativity, in our opinion our work is relevant for
the scope of this special issue on Extreme Regimes of Classical
and Quantum Gravity Models, bearing also in mind the relevance
of the BMS group for modern studies of black holes
\cite{BMS01,BMS03,BMS04}. Moreover, we possibly fill a gap
in the literature, because we have not found previous 
papers on the BMS group among those published in Symmetry. 
The original contributions of our paper are as follows. 
\vskip 0.3cm
\noindent
(i) Proof that to each normal elliptic transformation of
the complex variable $\zeta$ used in the metric for cuts of
null infinity there corresponds a BMS supertranslation. Although
this might be seen as a corollary of the work initiated in
Ref. \cite{BMS13}, it has prepared the ground for the items below.
\vskip 0.3cm
\noindent
(ii) Study of the conformal factor in the BMS transformation of
the $u$ variable as a function of the squared modulus of $\zeta$.
In the loxodromic and hyperbolic cases, such a conformal factor 
turns out to be either monotonically increasing or monotonically
decreasing as a function of the real variable given by the absolute
value of $\zeta$. In the parabolic case, the conformal factor is
instead a real-valued function of complex variable, and one needs
the plots of Figure $2$.
\vskip 0.3cm
\noindent
(iii) A classification of Killing vector fields of the Bondi metric 
has been obtained in Sec. $5$.
\vskip 0.3cm
\noindent
(iv) In Sec. $6$ we have found that BMS transformations are the
restriction to a pair of unit circles of a more general set of
transformations. Within this broader framework, the geometry of 
such transformations is studied by means of its Segre manifold.
This provides an unforeseen bridge between the language of
algebraic geometry and the analysis of BMS transformations in
general relativity.
\vskip 0.3cm
\noindent
(v) Our remarks at the end of Sec. $5$ might lead to a systematic
application of projective geometry techniques for the definition
of points at infinity in general relativity.
\vskip 0.3cm
\noindent
(vi) Our results in Sec. $5$ suggest four sets of Killing fields associated with the four 
branches of BMS transformations. As discussed in Sec. $3$, the elliptic transformations 
(the case with $K_\Lambda(\zeta)=1$) define the Abelian subgroup of supertranslations. 
The linearized action of supertranslations in the Schwarzschild case is already studed in 
\cite{BMS03} which results in a black hole with linearized supertranslation hair. It would be 
interesting to study the action of parabolic, hyperbolic and loxodromic transformations 
defined by the Killing fields (\ref{(5.12)}) on a black hole metric. 

To sum up, we have addressed the physical problem of obtaining
a more complete understanding of BMS diffeomorphisms of an asymptotically
flat spacetime. The tools we have developed might therefore lead to new
developments in black hole physics (see item (vi) above) and in the area
of geometric methods in theories of gravity, especially in light of the
original framework described in Sec. $6$.

\vskip 1cm

\renewcommand{\theequation}{A.\arabic{equation}}

\centerline {\bf Appendix A: composition of BMS transformations}
\setcounter{equation}{0}

\vskip 1cm

It is helpful to derive, with the notation in our section $2$, the
composition rule of two BMS transformations. For this purpose we
note that, since a BMS transformation yields
\begin{equation}
T(\zeta,u)=(\zeta',u')=(T(\zeta),T(u)),
\label{(A1)}
\end{equation}
where
\begin{equation}
\zeta'=T(\zeta)={(a\zeta+b)\over (c\zeta+d)}=f_{\Lambda}(\zeta),
\label{(A2)}
\end{equation}
\begin{equation}
u'=T(u)=K_{\Lambda}(\zeta)\Bigr[u+\alpha(\zeta,{\overline \zeta})\Bigr],
\label{(A3)}
\end{equation}
the subsequent BMS map leads to
\begin{equation}
T(\zeta',u')=(\zeta'',u'')=(T(\zeta'),T(u')),
\label{(A4)}
\end{equation}
where, by virtue of Eq. (A2), one obtains
\begin{eqnarray}
\zeta'' &=& T(\zeta')={(a'\zeta'+b')\over (c'\zeta'+d')}=f_{\Lambda'}(\zeta')
\nonumber \\
&=& {{(a'a+b'c)\zeta+(a'b+b'd)}\over 
{(c'a+d'c)\zeta+(c'b+d'd)}}={(A\zeta+B)\over (C\zeta+D)}
=f_{\Lambda''}(\zeta),
\label{(A5)}
\end{eqnarray}
having defined
\begin{equation}
\Lambda''=\left(\begin{matrix}
A & B \cr
C & D
\end{matrix}\right)
=\Lambda' \Lambda ,
\label{(A6)}
\end{equation}
which is the product of the ${\rm PSL}(2,\C)$ matrices
$$
\Lambda'=\left(\begin{matrix}
a' & b' \cr c' & d' 
\end{matrix}\right), \;
\Lambda=\left(\begin{matrix}
a & b \cr c & d 
\end{matrix}\right).
$$
Moreover, one finds
\begin{equation}
u'' = T(u')=K_{\Lambda'}(\zeta')\Bigr[u'
+\alpha(\zeta',{\overline \zeta}')\Bigr]
= {\widetilde K}_{\Lambda'\Lambda}
\Bigr[u+{\widetilde \alpha}(\zeta,{\overline \zeta})\Bigr],
\end{equation}
having defined
\begin{equation}
{\widetilde K}_{\Lambda' \Lambda}(\zeta) \equiv 
\Bigr[K_{\Lambda'}(f_{\Lambda}(\zeta))\Bigr]K_{\Lambda}(\zeta),
\label{(A8)}
\end{equation}
\begin{equation}
{\widetilde \alpha}(\zeta,{\overline \zeta}) \equiv
\alpha(\zeta,{\overline \zeta})
+{\alpha \Bigr(f_{\Lambda}(\zeta),{\overline f}_{\Lambda}(\zeta)\Bigr)
\over K_{\Lambda}(\zeta)}.
\label{(A9)}
\end{equation}

\vskip 100cm

\renewcommand{\theequation}{B.\arabic{equation}}
\centerline {\bf Appendix B: Origin and properties of fractional linear maps}
\setcounter{equation}{0}

\vskip 1cm
 
Suppose that the pair of complex coordinates $(z_{0},z_{1})$ are mapped
into the pair $(z_{0}',z_{1}')$ by the linear transformation
\begin{equation}
\left(\begin{matrix}
z_{0}' \cr z_{1}'
\end{matrix}\right)
= \left(\begin{matrix}
a & b \cr
c & d 
\end{matrix}\right)
\left(\begin{matrix}
z_{0} \cr z_{1} 
\end{matrix}\right), \; ad-bc \not = 0.
\label{(B1)}
\end{equation}
This means that the ratio $\zeta={z_{0}\over z_{1}}$ is mapped into
\begin{equation}
\zeta'={z_{0}'\over z_{1}'}={(az_{0}+bz_{1}) \over (cz_{0}+dz_{1})}
={(a\zeta+b)\over (c\zeta+d)}.
\label{(B2)}
\end{equation}
Thus, a fractional linear map arises from a linear transformation acting
on the homogeneous coordinates $z_{0},z_{1}$. For further insight, we
refer the reader to the lecture notes in Ref. \cite{BMS20}.

If the matrix on the right-hand side of Eq. (B1) pertains to
${\rm PSL}(2,\C)$, the condition of unit determinant yields
\begin{equation}
b={(ad-1)\over c},
\label{(B3)}
\end{equation}
and hence one finds \cite{BMS21}
\begin{equation}
\zeta'={(a\zeta+b)\over (c\zeta+d)}
={{{a \over c}(c\zeta+d)-{1 \over c}}\over (c\zeta+d)}
={a \over c}-{1 \over |c|^{2}}
\left({|c|\over c}\right)^{2}
{1 \over \left(\zeta+{d \over c}\right)}.
\label{(B4)}
\end{equation}
Thus, half of the BMS transformations as in Eq. (A2) arise by
composition of the following maps \cite{BMS21}:
\begin{equation}
{\rm Translation} \; \; \zeta \rightarrow \zeta+{d \over c},
\label{(B5)}
\end{equation}
\begin{equation}
{\rm Inversion} \; \; \zeta+{d \over c} \rightarrow
{1 \over \left(\zeta+{d \over c}\right)},
\label{(B6)}
\end{equation}
\begin{equation}
{\rm Rotation} \; \; \zeta \rightarrow - \left({|c|\over c}\right)^{2}z,
\label{(B7)}
\end{equation}
\begin{equation}
{\rm Dilation} \; \; \zeta \rightarrow 
{1 \over |c|^{2}}\zeta,
\label{(B8)}
\end{equation}
and eventually a further translation
\begin{equation}
\zeta \rightarrow \zeta+{a \over c}.
\label{(B9)}
\end{equation}

The interplay of homogeneous and non-homogeneous coordinates has not
been fully exploited in general relativity so far, to the best of our
knowledge.  For example, linear transformations among real homogeneous
coordinates may be a powerful tool for studying points at infinity.
In particular, one could imagine that the coordinates
$x^{1},x^{2},x^{3},x^{4}$ used for a real four-dimensional Lorentzian
spacetime manifold arise from five homogeneous coordinates
$(y^{0},y^{1},y^{2},y^{3},y^{4})$ according to the rule
\begin{equation}
x^{1}={y^{1}\over y^{0}}, \;
x^{2}={y^{2}\over y^{0}}, \;
x^{3}={y^{3}\over y^{0}}, \;
x^{4}={y^{4}\over y^{0}},
\label{(B.10)}
\end{equation}
the $y$'s being subject to the linear transformations
\begin{equation}
{y'}^{\mu}=\sum_{\nu=0}^{4}A_{\; \nu}^{\mu} \; y^{\nu}, \;
{\rm det}\left(A_{\; \nu}^{\mu}\right) \not =0,
\label{(B11)}
\end{equation}
which imply the following transformation rules for spacetime
coordinates:
\begin{equation}
{x'}^{\mu}={\sum_{\lambda=0}^{4}A_{\; \lambda}^{\mu} \; y^{\lambda}
\over \sum_{\nu=0}^{4}A_{\; \nu}^{0} \; y^{\nu}} \;
\forall \mu=1,2,3,4.
\label{(B12)}
\end{equation}
The equations (B12) might provide a fully projective way of studying
the concept of infinity (cf. Ref. \cite{BMS22}).

\end{document}